\documentclass{bvm2022} 
\makeatletter
\let\blx@rerun@biber\relax
\makeatother

\begin{document}
\selectlanguage{english} % Englisch

% Angabe des Titels Ihres Beitrages
\title{Initial Investigations Towards Non-invasive Monitoring of Chronic Wound Healing Using Deep Learning and Ultrasound Imaging}

\titlerunning{Monitoring of Chronic Wounds in US Imaging}

\author{
	Maja \lname{Schlereth} \inst{1,2}, 
	Daniel \lname{Stromer} \inst{2}, 
	Yash \lname{Mantri} \inst{3}, 
	Jason \lname{Tsujimoto} \inst{3}, 
	Katharina \lname{Breininger} \inst{1}, 
	Andreas \lname{Maier} \inst{2},
	Caesar \lname{Anderson} \inst{4},
	Pranav~S. \lname{Garimella} \inst{5},
	Jesse~V. \lname{Jokerst} \inst{6}
}

\authorrunning{Schlereth et al.}

\institute{
\inst{1} Department Artificial Intelligence in Biomedical Engineering, \\FAU Erlangen-N\"urnberg, Erlangen\\
\inst{2} Pattern Recognition Lab, FAU Erlangen-N\"urnberg, Erlangen\\
\inst{3} Department of Bioengineering, University of California, San Diego\\
\inst{4} Department of Emergency Medicine, San Diego\\
\inst{5} Division of Nephrology and Hypertension, Department of Medicine, San Diego\\
\inst{6} Department of Nanoengineering, University of California, San Diego
}

\email{maja.schlereth@fau.de}

\maketitle

\begin{abstract}
	Chronic wounds including diabetic and arterial/venous insufficiency injuries have become a major burden for healthcare systems worldwide. Demographic changes suggest that wound care will play an even bigger role in the coming decades. 
	Predicting and monitoring response to therapy in wound care is currently largely based on visual inspection with little information on the underlying tissue. Thus, there is an urgent unmet need for innovative approaches that facilitate personalized diagnostics and treatments at the point-of-care. It has been recently shown that ultrasound imaging can monitor response to therapy in wound care, but this work required onerous manual image annotations. 
	In this study we present initial results of a deep learning-based automatic segmentation of cross-sectional wound size in ultrasound images and identify requirements and challenges for future research on this application. 
	Evaluation of the segmentation results underscores the potential of the proposed deep learning approach to complement non-invasive imaging with Dice scores of $0.34$ (U-Net, FCN) and $0.27$ (ResNet-U-Net) but also highlights the need for improving robustness further. We conclude that deep learning-supported analysis of non-invasive ultrasound images is a promising area of research to automatically extract cross-sectional wound size and depth information with potential value in monitoring response to therapy.
\end{abstract}

\section{Introduction}
Chronic wounds affect around 6.6 million United States citizens per year (prevalence of ${\sim}2\ts\%$)~\cite{ref_1}. These wounds often lead to patient immobility, increased risk of sepsis and amputation, pain, decreased quality of life, and a shorter life expectancy. The five-year survival rate of patients with chronic wounds is significantly lower than for age- and sex-matched controls~\cite{ref_2}. Regular assessment of chronic wounds includes physical and visual examination or manual probing for tunneling wounds~\cite{ref_3}. Thus, these examinations mostly evaluate the skin surface. Tracking the kinetics of wound healing - especially below the skin surface - is still largely based on clinical experience. Using only surface information misses key information about the interface between healthy and diseased tissue and vasculature below the skin surface~\cite{ref_4}.
Assessment of cross-sectional wound size development can help to detect stagnant healing earlier. Thus, extracellular matrices or skin grafts can be applied earlier to improve the curative effects. Mantri et\,al.\ showed that non-invasive imaging such as ultrasound (US) imaging can reveal further insights into the healing processes by generating temporal and spatial information~\cite{ref_5}. 
US imaging is particularly well suited for wound care because it is affordable and already broadly used in various areas of medicine. However, it can also suffer from low contrast and imaging artifacts. Image interpretation can be strongly user-dependent; thus, there is a clear need for automatic assessment of wound size to increase the clinical value of US.
Wang et\,al.\ worked on a fully automatic segmentation of wound areas in photographic images~\cite{ref_6}. With this method, they are able to monitor superficial wounds but could not image the three-dimensional wound architecture below the skin surface. 
Huang et\,al.\ recently reviewed machine learning approaches applied on different modalities to objectively assess burn wound severity compared to subjective methods~\cite{ref_7}.
In this study, initial results for automatic segmentation of US images are presented. We evaluated three different deep learning segmentation networks in the context of US-based wound sizing: a fully convolutional network (FCN) with a DenseNet-like encoder, a ResNet-U-Net, and a U-Net architecture~\cite{ref_8,ref_9}. We also investigated the behavior of US intensities in different sections of the wound. Based on our evaluations, we identify challenges and define requirements for further research in this field to help physicians make objective wound assessments, support treatment decisions, and ultimately improve patient quality of life.

\section{Material and methods}
\subsection{Dataset}
All human subjects research was done with approval from the Internal Review Board at the University of California, San Diego. We used a commercially available LED-based photoacoustic/US imaging system (AcousticX from Cyberdyne Inc., Tsukuba, Japan). In this study, we used only the ultrasound mode and all US images were acquired at 30 frames/s. Details about the inclusion and exclusion criteria, the imaging equipment used, and manual image processing can be found in our prior work~\cite{ref_5}. We imaged 62 patients, affected by different types of chronic leg wounds. Subjects were scanned over a four-month period for a total of 161 scans containing 15,000 images. All scans included wound and healthy adjacent tissue. Ground truth (GT) annotations were done manually in ImageJ by outlining the border between wound and healthy tissue. Each image in a sweep scan was processed individually. Ultrasound imaging and annotations of the scans were performed by two experts in this field. Each scan was annotated by one expert due to the large annotation effort. For development of the networks, the dataset was divided into training, validation and test set with the data split shown in Table~\ref{figure_ucsd_data_dist}. 

\begin{table}[t]
	\caption{Distribution of US data for training, validating and testing.}
	\begin{tabular*}{\textwidth}{l@{\extracolsep\fill}lll}
		\hline
		& Training Set & Validation Set & Test Set \\
		\hline
		Number of Patients & 37 & 15 & 10 \\
		Number of Scans & 78 & 32 & 51 \\
		Number of Images & 7148 & 3061 & 5052 \\
		\hline
	\end{tabular*}	
	
	\label{figure_ucsd_data_dist}
\end{table}

\subsection{Deep learning networks and training strategy}
The three deep learning networks used for this work are FCN with a DenseNet encoder, ResNet-U-Net with a ResNet18 encoder and an adapted four-level U-Net with LeakyReLUs instead of standard ReLU activation functions. For DenseNet-FCN and the ResNet-U-Net, weights pre-trained on the ImageNet data set were used. For all networks, image input sizes of $320\times320\ts$px, Adam-Optimizer, batch size $3$ and decaying learning rate (start: 1$e$\textsuperscript{-3}, $\gamma$ for weight decay: $0.1$, step size: $10$ epochs) were used. All networks were implemented in Python using PyTorch and trained until convergence on an Nvidia GeForce GTX 1080. Hyperparameters were selected based on the validation set. 

To improve distribution of different wound type images in the training, validation, and test datasets, manifold learning-based data selection proposed by Chen was applied~\cite{ref_10}. As many images for training are heavily correlated, random augmentation techniques were introduced to make the training more robust (i.\,e.: random brightness, noise, saturation, rotation, left-right flipping)~\cite{ref_11}, and the images were normalized to match ImageNet statistics.

\subsection{Evaluation metrics} 
For quantitative evaluation of the network, we calculated precision, recall and the Dice score, also known as F1 score, of the network prediction compared to the GT annotation. Results are also visualized to compare GT with the network prediction. True positive (TP) pixels are colored green, false positives (FP) are colored yellow, false negatives (FN) are colored red and true negative (TN) areas are transparent. 

\subsection{Wound intensity evaluation}
\begin{figure}[b]
	\centering
	\setlength{\figwidth}{0.9\textwidth}
	\begin{subfigure}{\figwidth}
	    \includegraphics[width=\figwidth]{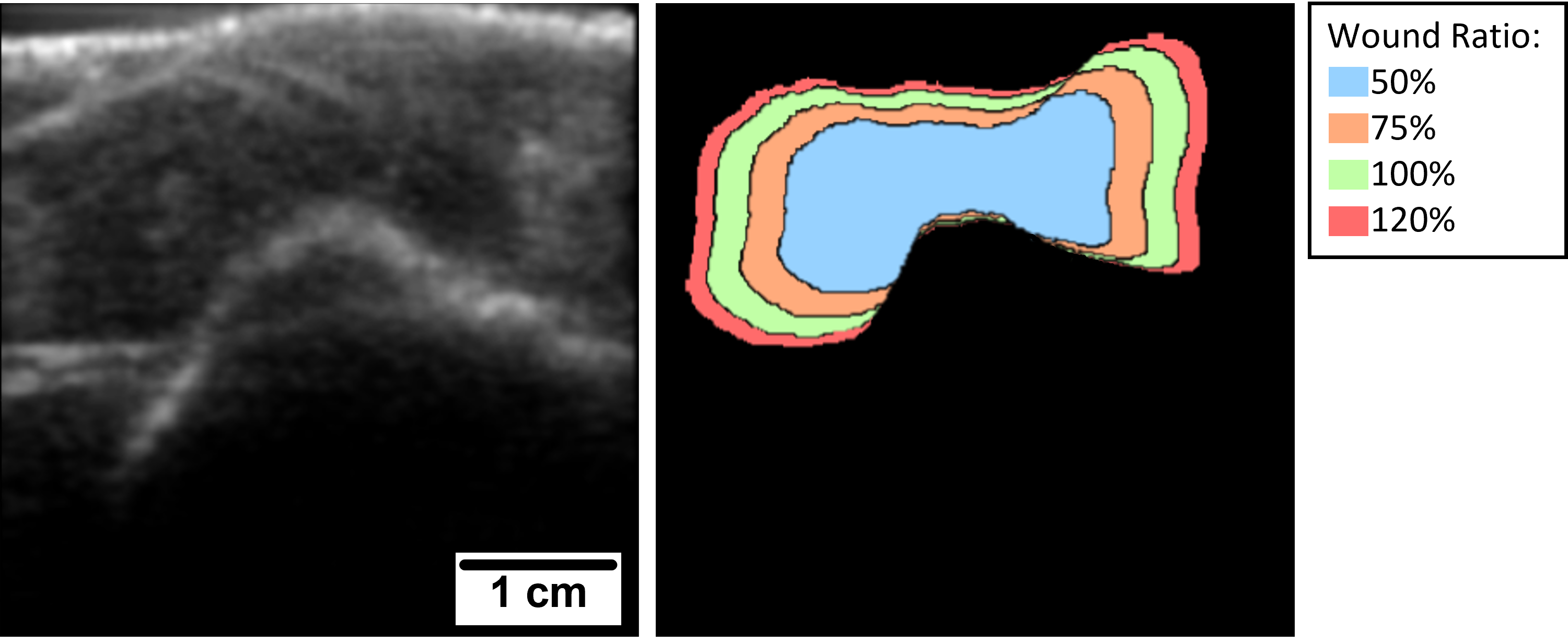}
	\end{subfigure}
	\caption{Different wound regions for comparison of US intensities $\overline{m}_{r}$ within the wound.}
	\label{figure_us_woundsize}
\end{figure}

To better understand the wound morphology and how its composition relates to measured US intensities, we analyze the average US intensity in different sections of the wound for all scans of the test set. The wound area is given by the GT segmentation mask for each scan. We introduce an evaluation scheme that adapts the segmented wound area by a dilation and erosion process. We calculated the ratio of the mean US intensity $\overline{m}_{r}$ of different wound regions compared to $\overline{m}_{w}$ over the whole wound. We used the regions 0--50$\ts$\% (blue), 50--75$\ts$\% (orange), 75--100$\ts$\% (green) and 100--120$\ts$\% (red) of the actual wound area. Figure \ref{figure_us_woundsize} shows the different wound regions.

\section{Results}
\begin{figure}[b]
    \centering
	\setlength{\figwidth}{0.22\textwidth}
	
	\begin{subfigure}{\figwidth}
    	\includegraphics[width=\textwidth]{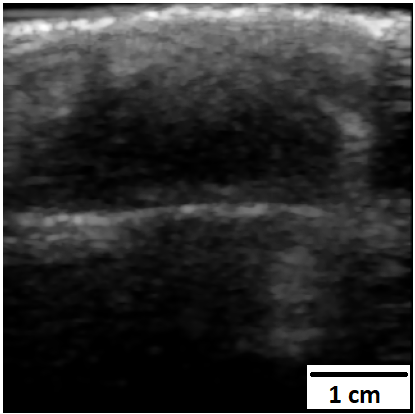}
    \end{subfigure}
    \hfill
    \begin{subfigure}{\figwidth}
    	\includegraphics[width=\textwidth]{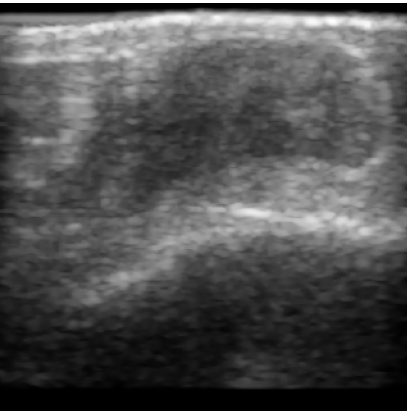}
    \end{subfigure}
    \hfill
    \begin{subfigure}{\figwidth}
    	\includegraphics[width=\textwidth]{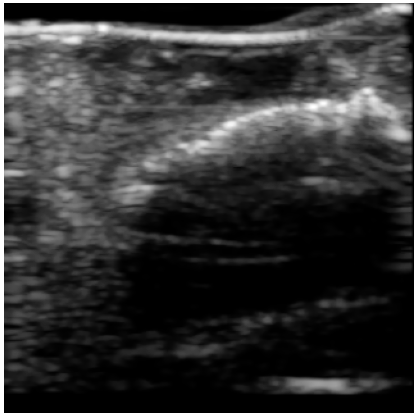}
    \end{subfigure}
    \hfill
    \begin{subfigure}{\figwidth}
    	\includegraphics[width=\textwidth]{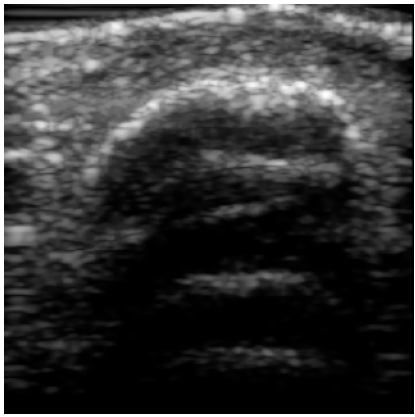}
    \end{subfigure}
    \medskip
    \begin{subfigure}{\figwidth}
    	\includegraphics[width=\textwidth]{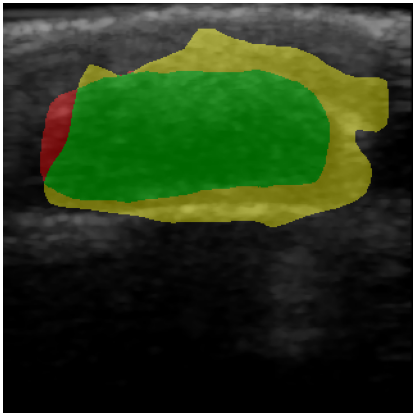}
    \end{subfigure}
    \hfill
    \begin{subfigure}{\figwidth}
    	\includegraphics[width=\textwidth]{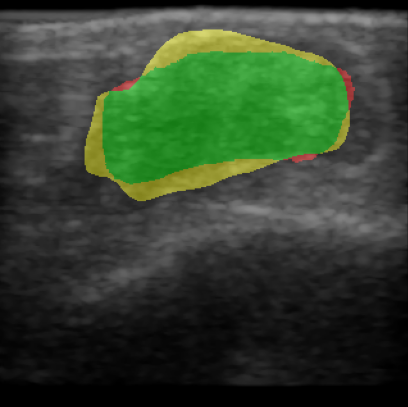}
    \end{subfigure}
    \hfill
    \begin{subfigure}{\figwidth}
    	\includegraphics[width=\textwidth]{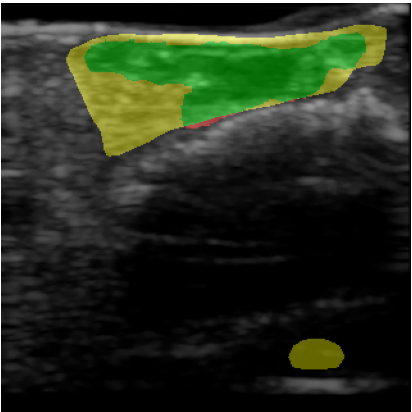}
    \end{subfigure}
    \hfill
    \begin{subfigure}{\figwidth}
    	\includegraphics[width=\textwidth]{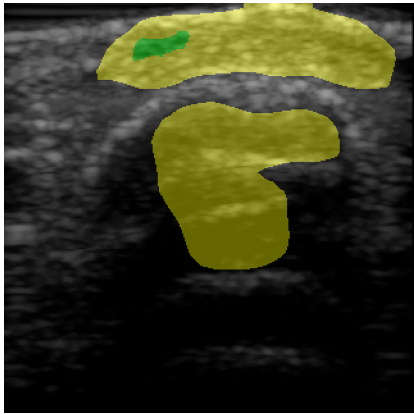}
    \end{subfigure}
    
	\caption{(a-d) Four representative US wound images. (e-h) Visualization of corresponding segmentation predictions for FCN. Green indicates correct matches, red and yellow indicate missed (false negative) and oversegmented (false positive) areas, respectively.}
	\label{figure_wound_visual}
\end{figure}

\begin{table}[t]
	\caption{Dice score, precision and recall value (mean $\pm$ standard deviation over all frames) of all trained networks.}
	\begin{tabular*}{\textwidth}{l@{\extracolsep\fill}lll}
		\hline
		Network & FCN & ResNet-U-Net & U-Net \\
		\hline
		Dice score & 0.34 $\pm$ 0.31 & 0.27 $\pm$ 0.32 & 0.34 $\pm$ 0.31 \\
		Precision & 0.27 $\pm$ 0.27 & 0.33 $\pm$ 0.40 & 0.33 $\pm$ 0.33 \\
		Recall & 0.71 $\pm$ 0.41 & 0.35 $\pm$ 0.37 & 0.55 $\pm$ 0.39 \\
		\hline
	\end{tabular*}
	\label{figure_f1scores}
\end{table}

\begin{table}[t]
	\caption{Comparison between US intensities measured in different ratios of wound size compared to 100$\ts$\% of actual wound size.}
	\begin{tabular*}{\textwidth}{l@{\extracolsep\fill}llll}
		\hline
		Wound Region & 0--50\ts\%  &50--75\ts\% & 75--100\ts\% & 100--120\ts\% \\
		\hline
		US Intensity Ratio& 0.89 $\pm$ 0.11 & 1.06 $\pm$ 0.08 & 1.24 $\pm$ 0.21 & 1.47 $\pm$ 0.37\\
		\hline
	\end{tabular*}
	\label{figure_comp_ratios}
\end{table}

Table~\ref{figure_f1scores} shows the quantitative evaluation of the algorithms. U-Net and FCN both achieve a Dice score of $0.34$ compared to pre-trained ResNet-U-Net which has a Dice score of $0.27$. Visualizations of the segmentation results for FCN are shown in Figure~\ref{figure_wound_visual}, where (a-d) are the original inputs, and (e-h) the corresponding color-coded outcomes. In general, the performance is promising for a number of cases, although we see a high FP rate (d,h) for others. 
Table~\ref{figure_comp_ratios} shows the ratio of $\overline{m}_{r}$ for different wound regions compared to $\overline{m}_{w}$ of the whole wound area. In the wound center, the US values are lower (ratio: 0.89 $\pm$ 0.11), compared to the wound borders (ratio: 1.47 $\pm$ 0.37).

\section{Discussion}
In this work, we showed a proof-of-concept for a non-invasive imaging technique paired with machine learning. We performed initial experiments for automatic segmentation of US wound images with deep learning which show an Dice score of $0.34$ for both U-Net and FCN and $0.27$ for ResNet-U-Net. The segmentation results strongly depend on the quality of the scan and the specific wound type. Quantitative results can still be improved but the visual examination of the results indicates applicability of the proposed setup. For images illustrated in Figure~\ref{figure_wound_visual}(e-g), the segmentation is very good. However, the network misclassifies the region beneath the hyperechoic bone surface (image h) as this region therefore appears dark and similar in intensity to the wound region.

The evaluation of the mean US intensity ratio in different wound regions show higher ratios for wound borders and lower intensities for the wound center compared to the whole wound area. Low echoic region can indicate the absence of healing  tissue. Higher US intensity ratios at the wound edges could imply hyperdense or scar tissue. Generally, an increase in echogenicity indicates tissue regeneration and wound contraction~\cite{ref_12}. This information about wound characteristics can help with automatic wound assessment in future work.

Based on our results, we identify the following challenges: US images are notoriously difficult to interpret because they can have low contrast as well as imaging artifacts. Given the signal variability of wound tissue observed in US images, larger amounts of training data are needed to allow for a robust determination of wound size and healing progress. While a prior study showed no significant inter-observer bias for annotating wound areas in US images~\cite{ref_5}, a study to better understand inter-observer variability from different institutes and to overcome the subjective perception of a single expert during the annotation process is needed. To facilitate this, a standardized measurement protocol and a detailed annotation process can help to make scans more comparable. As a result, we expect an improved performance for a machine-learning based analysis by either increasing label consistency or by being able to explicitly model annotation uncertainty during the training. 

Along this path, we see the potential for automatic scoring and classification of wounds into different severity stages and a continuous non-invasive monitoring of therapy response by combining US imaging and machine learning, which can be implemented at the point-of-care for wound assessment. In this context, this monitoring can be integrated in eHealth applications, thus enabling improved personalized healthcare and therapy outcomes.

\begin{acknowledgement}
	JVJ acknowledges NIH support under grant R21 AG065776. 
\end{acknowledgement}

\printbibliography

\end{document}